# The use of Reverse Micelles in Downstream Processing of Biotechnological Products

Kai Lun LEE, CID:00664757      Ching Chieng CHONG, CID:00534092

*Department of Chemical Engineering and Chemical Technology,
Imperial College London, March 2011*

**Abstract**

This paper aims to discuss the use of reverse micelles in downstream processing of biotechnological products. The interest in this technology is piqued firstly by various advantages of a liquid-liquid extraction process, among which are cost effectiveness, and ease of scaling up and implementing a continuous process for whole broth processing. The use of reverse micelles is thought to be among the most promising due to the high efficiency and selectivity being achieved in some systems. However, there are various issues that have impeded the widespread use of reverse micelles such as the identification and development of suitable surfactants and ligands; as well as difficulties in the back extraction process. These issues, as well as latest developments and applications of reverse micelles in downstream processing of biotechnological products will be discussed in this paper.

## INTRODUCTION

Downstream processing of biotech products consists of different unit processes which depend on the characteristics of the product, the required level of purification and whether the product is intracellular or extracellular. If biological cells are not involved in the production, the fermentation broth can be directly concentrated and then purified to get the final form of product; otherwise, the products can be classified as either intracellular (formed inside the cell) or extracellular (secreted into the surrounding medium). The different stages involved in downstream processing of products involving biological cells are shown in the figure to the right.

The first stage involves the separation of cells from the fermentation broth. If the product is intra-cellular, the cell has to be disrupted to release the product contained within. The main difference between product concentration and purification is that the latter involves separation of unwanted components which closely resemble the product in its physical and chemical form. Product purification is also the most expensive stage. The final stage is product polishing which is to pack the purified product into a form which is stable, easily portable and convenient to use. Unlike usual chemical engineering separation processes, quality is more important than quantity for downstream processing in the biotech industry.

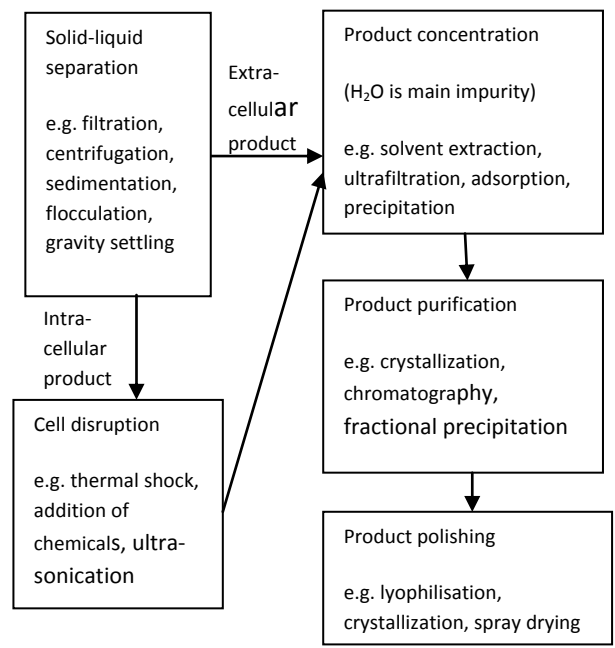

Figure 1: Overview of the stages involved in downstream separation processing

Reverse Micelles in Downstream Processing

The selective separation and purification of target proteins in a mixture of both similar and dissimilar proteins, as well as other biological and chemical compounds is one of the most common and important processes and this paper focuses on the use of reverse micelles (RMs) in this separation. RMs are thermodynamically stable, nanometer-sized assemblies of surfactants that encapsulate microscopic pools of water in a bulk organic phase. This allows proteins and other hydrophilic molecules to be solubilised in the aqueous microenvironment while organic reactants and products remain in the bulk organic phase. RMs are dynamic quantities, colliding with each other in solution and occasionally exchanging contents. Approximately one collision in one thousand results in an exchange of RM contents (Fletcher et al., 1987). Collisions occur on a timescale of nanoseconds while exchanges of content occur every few microseconds.

The possibility of using RMs to solubilise proteins for protein separation was first proposed by Luisi et al. (1979). The application of RMs for bioseparation has attracted considerable attention in the past three decades because it is considered to have high potential for the large-scale downstream separation of biomolecules from fermentation mixtures (Liu et al., 2008).

**BACKGROUND**

The phase-transfer between bulk aqueous and surfactant-containing organic phases is the basis for extraction of proteins from aqueous solutions. It is as shown in **Figure** .

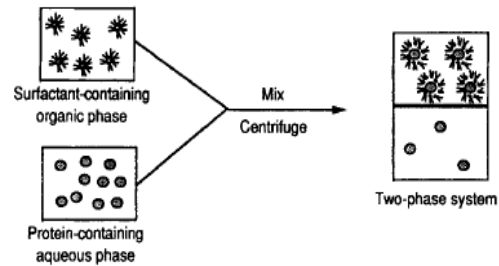

Figure 2: Phase transfer (Matzke et al., 1992)

There are two structural parameters associated with RMs which are the water content of the RMs ($W_0$) and the aggregation number ($N_{ag}$). Their definitions are given by the following equations.

(Hai & Kong, 2008): $$W_0 = \frac{[\text{water}]}{[\text{surfactant}]}$$

(Thévenot et al., 2005):
$$N_{ag} = \frac{\text{average molecular weight of RM}}{\text{molecular weight of the surfactant}}$$

$N_{ag}$ enables the number of micelles in a system to be calculated provided the surfactant concentration is known (Matzke et al., 1992) while $W_0$ relates to the size of the RM as shown below.

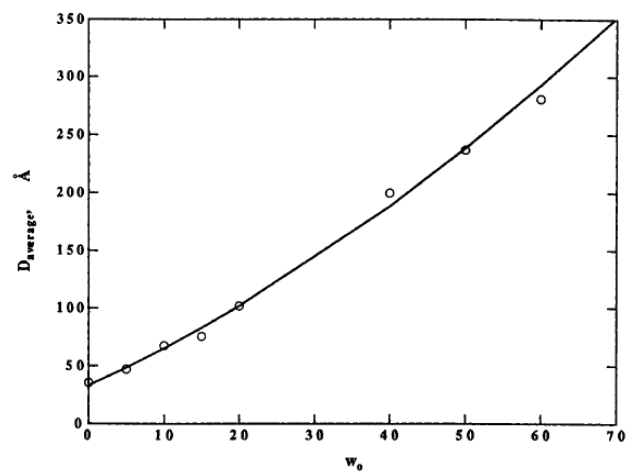

Figure 3: Average hydrodynamic diameters for AOT reverse micelles at different water contents

Three important questions raised by Luisi et al. (1988) are

(i) what are the driving forces for the solubilisation of protein;

(ii) what is the localization of proteins in the reverse micelle;

(iii) what size or shape perturbation is induced in the reverse micelle droplets by the protein solubilisation?

These questions were addressed when the water shell model was proposed by Bonner et al. (1980) as a mechanism for protein solubilisation. Various researches have been undertaken to investigate the factors affecting protein solubilisation by RMs have been carried out and their results are summarized in Table 1.

Table 1: Factors affecting protein solubilization by RMs

| Factor | RMs and proteins under investigation | Conclusion |
|---|---|---|
| **Water content of RM, $W_0$ and proteins** (Hai & Kong, 2008) | RM: Sodium bis(2-ethyhexyl) sulfosuccinate (AOT) Protein: BSA | There is a minimum $W_0$ required for the solubilization of a fixed amount of protein. The diameter and viscosity of AOT reverse micelle increases with increasing $W_0$. Incorporation of protein into the micelles does not affect their mean diameter significantly |
| **Aqueous phase pH and ionic strength** | RM: AOT Protein: BSA Salts: KCl, NaCl, $CaCl_2$ and $MgCl_2$ | Minimum $W_0$ is independent of pH for $MgCl_2$ whereas increasing the concentration of KCl and $CaCl_2$ decreases minimum $W_0$. The pH range at which maximum protein solubilization efficiency is attained is dependent on the type of salt dissolved in the aqueous phase. For NaCl and $CaCl_2$, increasing the salt concentration will increase the solubilization efficiency |
| **Surfactant concentration** (Shin & Vera, 2002) | RM: DODMAC Protein: lysozyme | The concentration of DODMAC has negligible influence on the amount of lysozyme extracted. The solubilization limit is strongly affected by the cations present in the aqueous solution but not affected by the temperature (room temperature to 35 °C). The presence of sodium ions resulted in greater extraction than potassium and calcium ions. Complete lysozyme removal from the aqueous phase was observed when the pH is one unit greater than the pI of lysozyme. There is a solubilization limit for lysozyme in the organic phase as not all of it is extracted into the RMs. There were some which precipitated on the aqueous-organic interface |

| **Co-surfactant type** (Lee et al., 2004) | RM: AOT | Back extraction was found to be dependent on the species and concentration of the alcohol and carboxylic acid added to the RMs. |
|---|---|---|
| | Proteins: β-lactoglobulin, CAB, lipase | |
| | Co-surfactants: alcohol, carboxylic acid | Co-surfactants that suppresses the formation of RM clusters enhances back-extraction |

Exerting a larger influence on the stability of the solubilised proteins in RMs than the aforementioned factors is the nature of the surfactant itself. Various reverse micellar systems could be derived from different surfactants molecules, as summarized in Table 2.

Ionic surfactants are most commonly used, due to their ability to solubilise a wide range of proteins. However, the strong electrostatic interactions between proteins and ionic surfactants interaction cause the denaturation of proteins and subsequently low yield of target proteins.

To avoid this, Sawada (2004) and Singh (2006) proposed and investigated the use of non-ionic surfactants. Limited success has been achieved with non-ionic surfactants due to difficulties in the extraction process, which results in similarly low yields despite the preventing denaturation. The selectivity of non-ionic RM processes are also low due to the weak forces, so similar proteins cannot be easily separated.

Mixed RMs with ionic and non-ionic surfactants were developed to address these issues by reducing the undesirable effects of either ionic or non-ioninc RMs alone. Results have shown that mixed RMs are more efficient at extracting and separating proteins than ionic RMs but have limited applications due to their complexities.

Affinity-based RMs are composed of surfactants (ionic or nonionic) coupled with affinity ligands. The introduction of affinity ligands allows enhanced selectivity and extraction capacity. These novel forms of RMs are further elaborated in a section below.

Table 2: Reverse micellar systems

| Reverse micellar system | Example of surfactant molecules in the system | Major interaction forces during extraction |
|---|---|---|
| **Ionic surfactant-based RMs** (Liu J., 2004) (Shen, 2005) | AOT; CTAB | electrostatic interactions |
| **Non-ionic surfactant-based RMs** (Sawada, 2004) (Ichikawa et al., 2000) (Singh, 2006) | Tween 85; phospholipids; TX-100 | hydrophobic and hydrogen-bonding interactions |
| **Mixed RMs** (Fan, Ouyang, Wu, & Lu, 2001) (Goto et al., 1998) (Shen, 2005) | AOT-Tween80; AOT-DOLPA; AOT-OPE4; CTAB-TRPO | electrostatic interactions dominate hydrophobic and hydrogen-bonding interactions |
| **Affinity-based RMs** (Liu et al., 2008) | Cibacron modified lecithin; antibodies-liganded AOT | affinity interaction between proteins and their affinity ligands |

Back Extraction Process

The use of RMs in purification of proteins involves two processes, the forward and backward extraction processes. In the former, protein is transferred from a bulk aqueous phase to the water pool of RMs in an organic phase; while in the latter these proteins are recovered from the RMs into a fresh aqueous phase.

There are two significant problems with the back extraction process, namely, a decrease in activity yields due to structural changes in proteins as a result of the strong interactions between proteins and micelles; and a slow rate of back extraction due to the greater interfacial resistance towards protein release at the oil-water interface during the back extraction.

The first problem can be circumvented by the use of non-ionic surfactants, of which affinity based surfactants are particularly promising, albeit not being sufficiently studied to be applied to a broad range of proteins commonly extracted.

Three notable techniques to improve the rate of back extraction are:

1) using aqueous stripping solution with high salt concentration or high pH, or varying the temperature of the system.
2) Addition of appropriate alcohol
3) Addition of destabilizing solvent (Woll et al., 1989)or dehydrating aqueous phase of RMs with silica (Leser et al., 1993) gel or molecular sieves (Ram et al., 1994).
4) Addition of suitable counterionic surfactant.

The first technique utilises the electrostatic repulsion that occurs between the surfactant and protein as a result of the pH difference or the size exclusion when salt concentration is increased. However, these harsh conditions of an extreme pH environment or high salt concentration may reintroduce the problems of reduced specific activity of the proteins (Kinugasa T, 1992).

The second strategy involves the addition of suitable alcohol species during the back extraction process. Daliya & Juang (2007) found that alcohols which did not have a marked influence on the structure of proteins promoted the back extraction process when added at the Critical Alcohol Concentration. These alcohol reduces the interfacial tension of the reverse micelles, and promotes their fusion at the organic and aqueous interface; causing the release of proteins into the aqueous phase, as shown in Figure 3.

The last technique mentioned involves the use of a counterion surfactant such as TOMAC or DTAB, which are oppositely charged to the surfactants. They cause a "rapid collapse of the RMs, and a sharp decrease of the water content in the organic phase", which results in the quick release of proteins into the aqueous phase (Jarudilokkul, Poppenborg, & Stuckey, 1999).

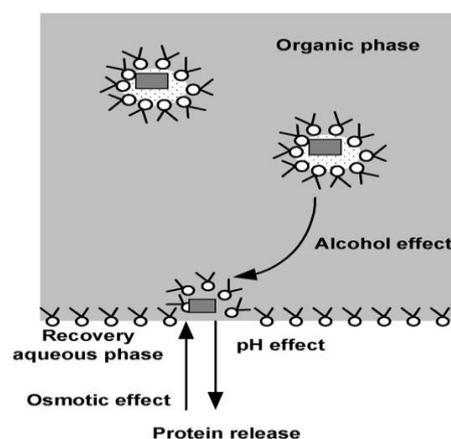
Figure 4 Effect of alcohol on back extraction process

## RECENT DEVELOPMENTS

Counter-Current Chromatography

Counter-current chromatography (CCC) has been reported as a very useful way for separation and purification of biopolymers such as proteins and nucleic acids. It was introduced in the 1970s and since then has gone through several development phases which have improved its design and efficiency. CCC involves liquid-liquid partition chromatography without any solid support matrix as the stationary phase is retained in the column using gravity or centrifugal force. This is beneficial because it allows elimination of all complications due to the solid support (Shibusawa & Ito, 1991). As most proteins are not soluble in the organic phase, protein separation via CCC uses two-phase aqueous systems.

Shen & Yu (2007) investigated the possibility of protein separation and enrichment by CCC using RM solvent systems. In order to carry out the separations, it was necessary to search for a two-phase solvent system that provides suitable partition coefficients (K) for target compounds. Since the major parameter for partition manipulation is the electrostatic interaction between the protein and charged heads of the ionic surfactant, they investigated the effect on protein separation efficiency when pH and ionic strength gradient were applied simultaneously. The RM system involved was AOT in n-hexane which makes up the stationary phase while the mobile phase was comprised of aqueous KCL.

Figure 5 shows the results obtained for separation using pH gradient elution. Myoglobin was essentially un-retained and was collected at early stage. However, it was cross-contaminated with cytochrome c and lysozyme which could be due to non-specific protein/micelle interactions. The total amount of lysozyme from fractions 14-20 was approximately 30% of the injected sample. Although these three proteins were not well resolved, the major portions of them emerged on the chromatogram in an order, as expected, essentially according to their pI values.

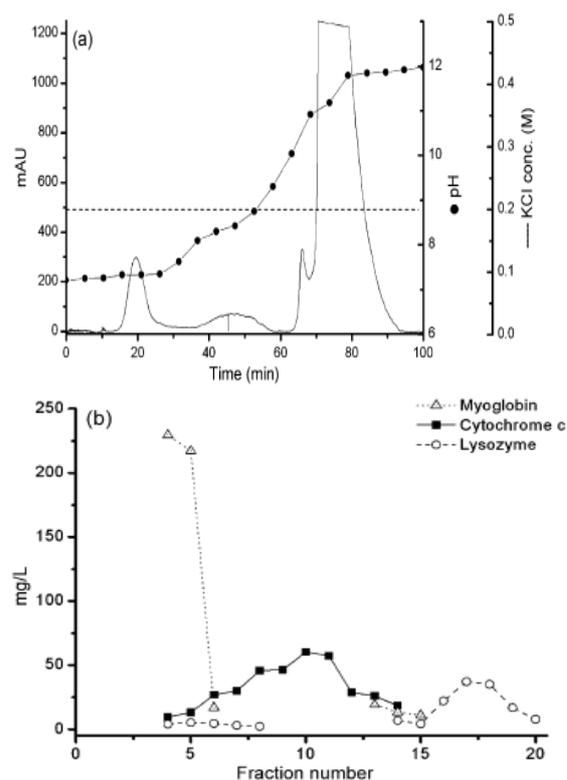

Figure 5: Separation using pH gradient elution with 0.2 M KCl. (a) Chromatogram obtained using on-line UV detector monitored at 280 nm, (b) protein concentrations of 20 fractions acquired by HPLC analysis

The results for separation using pH and KCl concentration gradient elution are illustrated in Figure 6. There was significantly less cross-contamination by the un-retained myoglobin and the separation of cytochrome and lysozyme was greatly enhanced. In addition, cytochrome recovery was 82% while lysozyme recovery was enhanced from 30% of the previous run to 90% in this experiment due to the high KCl concentration in the later stage of the elution.

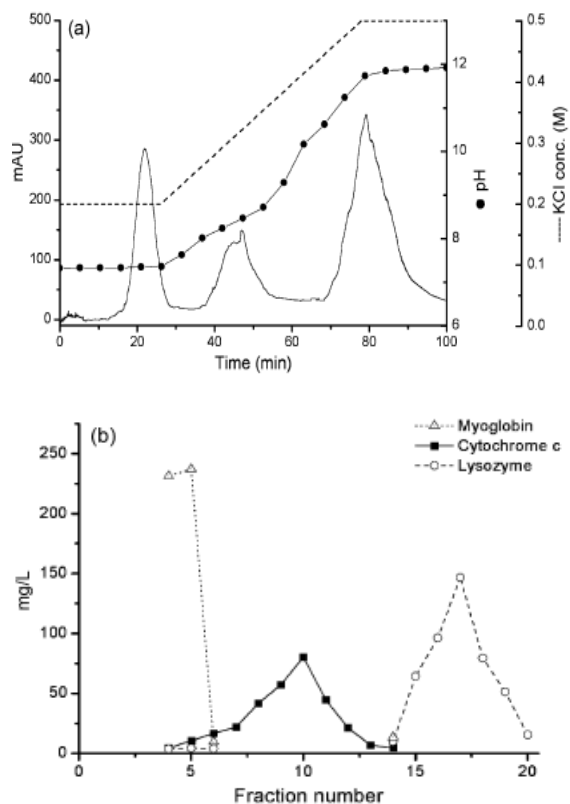

Figure 6: Separation using pH gradient KCL gradient elution, (a) Chromatogram obtained using on-line UV detector monitored at 280 nm, (b) protein concentrations of 20 fractions acquired by HPLC analysis

Enrichment of these two proteins was also achieved from a large-volume sample load (in a volume equal to the column capacity) showing that this technique could be potentially employed in the enrichment and recovery of proteins from large-volume aqueous solutions.

Affinity Based Reverse Micelles

Affinity-based Reverse Micelles Extraction and Separation (ARMES) have become of increasing interest of scholars due to the ability of ARMES to separate proteins with higher selectivity and higher purification levels as compared to ionic, non-ionic and mixed RMs, both of which are highly valued process qualities in downstream processing in the biotech industry as aforementioned. Comparison between conventional ionic RMs and ARMES in the table at the bottom of the page shows that typically ARMES techniques can produce a purification of an order of a magnitude higher.

ARMES can be broadly classified into two categories by their affinity ligands: i) Specific Ligands, with affinity ligands that have very narrow specificities for single compounds; and ii) Group Ligands, with affinity ligands capable of group specific interactions, to bind a range of similar compounds within "groups" (Liu et al., 2008).

The ARMES process involves four main steps:

i) Formation of a ligand-ligate complex
ii) Selective removal of the complex by reverse micelles
iii) Disassociation of complex into stripping solution
iv) Separation of ligand and ligate, and regeneration of the ligate.

The figure on the next page illustrates steps i and ii which make up the forward extraction process, and step iii in the back extraction process. It is imperative that the ligand is specific to the target ligate protein(s) and not to other undesired contaminants, forms a ligand-ligate complex that can be selectively removed from the mixture, and lastly, be easily separated from the ligate after disassociation (Paradkar & Dordick, 1993)

The extraction of the ligang-protein ligate complex into the reverse micellar organic phase in step ii is dependent on the system parameters, namely pH, ionic strength of aqueous phase, surfactant identity and surfactant concentration, as well as on the characteristics complex itself which includes effective pI of the complex, size of the complex and distribution of charge on the complex (Liu, Dong, & Sun, 2008).

Table 3: Comparison of Ionic and Affinity RMs

| Type | Reverse Micelle | Protein | Purification factor | Recovery | Reference |
|---|---|---|---|---|---|
| Ionic | AOT | nattokinase | 2.7 | 80% | (Liu et al., 2004) |
| | AOT | arginine deiminase | 4.52 | 85% | (Li et al., 2008) |
| Affinity | anti-CTN antibodies | lipase | 10.8 | | (Adachi et al., 2000) |
| | CB-Span 85 | lysozyme | 21.2 | 71% | (Liu et al., 2007) |

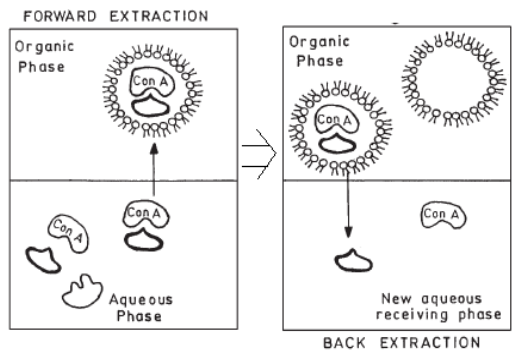

Figure 7: Illustration of the ARMES process

ARMES can be incorporated with both ionic and non-ionic RMs, though the former, AOT in particular, is much more common. This is in spite of strong electrostatic interactions in ionic surfactants impeding affinity under usual extractive conditions (Liu, Dong, & Sun, 2008). Adachi et al. (2000) and Sun et al. (1999) therefore proposed the incorporation of ARMES into non-ionic surfactants. This technique utilises the small extractive ability of non-ionic RMs such that only the affinity effect solubilises the target protein. However, the use of non-ionic ARMES is still limited due to the lack of suitable non-ionic surfactants available and also due to difficulties in phase separation i.e. separating such RMs from the clear interfaces (Adachi et al., 2000).

The most recent development in the work on incorporating ARMES in non-ionic RMs is the use of metal chelates which proved useful in enhancing the extraction of, or refolding proteins with histidine groups (or poly-histidine tagged). Such metal chelates are often incorporated using cosurfactants with a metal chelating hydrophilic head in addition to the non-ionic surfactant. Dong, Feng, & Sun (2009, 2010) successfully demonstrated the use of di(2-ethylhexyl) phosphoric acid (HDEHP) as a metal chelator in reverse micelles for the coupling of transition metal ions in the extraction of myoglobin from a mixed protein system; and later, the use of a Ni(II) chelator with non-ionic surfactants in the purification of recombinant his-tagged enhanced green fluorescent protein.

The high specificity and mild environment which preserves the native structure of proteins are distinguishing features of ARMES and may prove to be very useful industrially, in addition to the advantages provided by a liquid-liquid extraction process elaborated earlier. Despite not being sufficiently developed to be applicable to a large range of proteins, ARMES and in particular non-ionic or metal chelating ARMES, hold the potential to be the ideal protein separation technique, with high specificity and mild conditions perservering the native structure of proteins its distinguishing characteristics.

Enzymatic Reactions in RMs

The use of reverse micelles to solubilise enzymes in organic solutions is another recent development that has attracted interest from researchers. This is due to three main reasons; the first being that an aqueous environment can be created in RMs for hydrophilic enzymes which would be denatured in an organic phase, yet retain an organic phase for any lipophilic reactants or products (Stamatis, Xenakis, & Kolisis, 1999); a common example and area of research would be lipases and triacylglycerols since the solubility of the latter is largely improved in organic solvents. The second would be that such a system allows the use of surface active enzymes, i.e. enzymes that are active at the aqueous-organic interface. The third is that the aqueous-organic interface of such systems is usually very large (approximately $10^8$ $m^2/m^3$) due to the small size of the RMs (Verhaert & Hilhorst, 1991)

The microencapsulation of enzymes, including lipases, in RMs can be classified into three most common methods (Carvalho & Cabral, 2000):

i) Injection:
injecting concentrated aqueous enzyme solution into organic solution containing surfactant and mixing.

ii) Phase transfer:
phase transfer techniques between aqueous phase and organic solvent containing surfactant.

iii) Dissolution:
adding lyophilised enzyme to 2-phase RM solution already containing aqueous phase.

The injection method is the most common method used due to its simple procedure, while the phase transfer technique requires control of $W_0$ and is a length process, and the dissolution technique may cause deactivation of enzymes during the process (Carvalho & Cabral, 2000). The solubilisation of enzymes into RMs using these techniques depend on the pH and ionic strength of the aqueous phase, the sizes of the enzyme and RM, as well as the surfactant (Matzke et al., 1992). A comparison of these enzyme microencapsulation techniques by Matzk et al. (1992) shows that they result in different enzyme solubility and that while the dissolution technique required the micelle dimension to be similar or larger than the enzyme dimension, the injection method did not.

Membrane separations using reverse micelles in nearcritical and supercritical fluid solvents

Over the past two decades, there has been extensive research on the use of supercritical fluids for routine separations (Yonke et al., 2003). However, by nature, these separations are usually limited to batch processes and require a lot of energy to recycle the solvent to its original thermodynamic state for further processing.

Membrane separations in supercritical fluids, particularly supercritical $CO_2$ has been the focus of research over the last 5 to 7 years. By using this technique, the solute could be removed from the fluid solvent without a significant change in the thermodynamic state of the fluid. Although this has great potential to enhance supercritical fluid extraction and separation processes, it is still relatively limited to non-polar molecules (Yonker et al., 2003).

Yonker et al. (2003) investigated the use of RMs coupled with ultrafiltration for the separation of polar macromolecules dissolved in the cores of the RMs using nearcritical and supercritical fluid solvents. Separation occurs via a simple size exclusion process through the membrane pore, based on molecular weight. They demonstrated that neither dextran nor the protein were directly soluble in the pure fluids but were soluble in the aqueous RM core. Hence, this methodology extends the application of membrane separations in supercritical fluids to include both nonpolar and polar molecules.

However, the long-term exposure of the membrane to nearcritical and supercritical fluids was not investigated as the membrane separation process proceeded for 2 hours only. Furthermore, it is essential to have a good understanding of the membrane transport and separation process in both near-critical and supercritical fluid solvents in order to optimize the design of the membrane system.

**Conclusion**

Traditional separation processes such as electrophoresis or chromatography can become expensive and therefore economically unviable unless the product of interest is of high value. Hence, there is a need to develop cost-effective and efficient downstream processing methods.

Reverse micellar extraction is a promising liquid-liquid extraction technique that has the potential to be such a solution. Other factors making reverse micellar extraction an attractive area for research is that this technique is easy to scale-up and offers continuous operation. This technique has received immense attention, in particular for the isolation and purification of proteins and enzymes in recent times.

On the lab scale, reverse micellar extraction for downstream processing of proteins is well established but has not been reported to be used in industrial or pilot scale studies. The focus of research has been on the various factors affecting the efficiency of the extraction process such as surfactant and co-surfactant type, water content of the reverse micelles, pH, ionic strength and surfactant concentration.

This paper has discussed several recent developments in RM systems that have improved its potential to be industrially viable. The first discussed was the use of counter current chromatography which allows increases the ease of implementing a continuous system. The second discussed was the use of affininity based reverse micelles in separation and extraction (ARMES). The use of ARMES increases selectivity and purification, which are highly valued in biotechnological products. Additionally, ARMES potentially increases the number of systems where the use of non-ionic surfactants is possible, thereby preventing the denaturation of proteins during the extraction process which commonly occurs when ionic surfactants are used. The third area discussed was the implementation of enzymatic reaction in reverse micelles, which allows reactions between hydrophilic enzymes and both hydrophilic and hydrophobic reactants, and which provides a platform for surface active enzymes that may be used to create biopharmaceutical products. The fourth and final area discussed was the use of reverse micelles in supercritical or near-critical fluids, which while not fully investigated, potentially allows the separation and extraction of non-polar molecules in supercritical or near-critical fluids via size exclusion using membrane separation techniques.

While the use of reverse micellar systems have not been investigated industrially, the potential benefits and cost effectiveness of using such systems are great. With more research and development of these systems that allows application in a wider range of systems and proteins, as well as the solution of current problems with denaturation of proteins in ionic surfactant reverse micellar systems and back extraction issues; reverse micellar systems could well be the future de facto industry standard for downstream biotechnological separations of proteins.